# Effects of Photopatterning Conditions on Azimuthal Surface Anchoring Strength

Nilanthi P. Haputhanthrige [1,2], Mojtaba Rajabi [1], and Oleg D. Lavrentovich[1,2,3,4]

[1] *Advanced Materials and Liquid Crystal Institute, Kent State University, Kent, OH 44242, USA*

[2] *Department of Physics, Kent State University, Kent, OH 44242, USA*

[3] *Materials Science Graduate Program, Kent State University, Kent, OH 44242, USA*

[4] *Faculty of Chemistry, University of Warsaw, Zwirki i Wigury 101, 02-089 Warsaw, Poland*

**Correspondence**: Oleg D. Lavrentovich, olavrent@kent.edu

## Abstract

Spatially varying alignment of liquid crystals is essential for research and applications. One widely used method is based on the photopatterning of thin layers of azo-dye molecules, such as Brilliant Yellow (BY), that serve as an aligning substrate for a liquid crystal. In this study, we examine how photopatterning conditions, such as BY layer thickness ($b$), light intensity ($I$), irradiation dose, and age affect the alignment quality and the strength of the azimuthal surface anchoring. The azimuthal surface anchoring coefficient, $W$, is determined by analyzing the splitting of integer disclinations into half-integer disclinations at prepatterned substrates. The strongest anchoring is achieved for $b$ in the range of 5–8 nm. $W$ increases with the dose, and within the same dose, $W$ increases with $I$. Aging of a non-irradiated BY coating above 15 days reduces $W$. Our study also demonstrates that sealed photopatterned cells filled with a conventional nematic preserve their alignment quality for up to four weeks, after which time $W$ decreases. This work suggests the optimization pathways for photoalignment of nematic liquid crystals.

**Keywords:** Azimuthal anchoring, Surface anchoring, Liquid crystals, Photopatterning conditions, Alignment stability, Brilliant Yellow

## 1. Introduction

Alignment of liquid crystals (LCs) is crucial for their applications. While the unidirectional alignment can be achieved by mechanical rubbing [1-5], alignment with a spatially varying "easy axis" requires a more sophisticated approach. Spatially varying alignment becomes exceedingly important in many academic and applied projects, such as the fabrication of planar optics elements [6-12], LC elastomer coatings with predesigned topography [13-19], orientationally ordered environments that control collective and individual dynamics of microswimmers [20-25], and substrates that align living tissues [26-29]. The most popular approach to achieving the spatially varying alignment of LCs is photoalignment [13,30-38]. Photoalignment allows one to design complex director patterns with high spatial resolution [34,35,39], controllable surface anchoring [40], dynamic director alignment [41], and the capability to pattern the alignment on flexible and curved substrates [42-44]. Photoalignment does not induce impurities, electric charges, or mechanical damage to the treated surfaces, unlike conventional rubbing [45].

Brilliant Yellow is one of the azo-dye materials used for photoalignment [46-48]. Extensive studies by Yaroshchuk et al. [49] brought the conclusion that BY provides "excellent photoalignment of nematic, smectic ferroelectric and reactive liquid crystals" and "shows extraordinarily high photo and thermal stability". The study also noted that the double C=C bond in the core of BY molecules results in an absorption peak at 432 nm [49]. This peak, being in the lower part of the visible spectrum, might result in reduced stability when exposed to visible light. Nevertheless, the study by Yaroshchuk et al. [49] demonstrated that BY has excellent stability when irradiated with ultraviolet light and kept at an elevated temperature of 150 °C for one hour. Since BY allows one to produce high-quality alignment in a relatively easy process, it continues to attract interest in research and applications, see, for example, some recent publications in Refs. [33,49-54].

Azo-dye molecules undergo photoinduced reorientation when exposed to light [50,55,56]. For irradiation with a linearly polarized beam, the probability of absorption is $P \propto \cos^2\beta$ where $\beta$ is the angle between the long axis of the molecule in its *trans* state and the light polarization direction. The absorption–reorientation process repeats itself until the dye molecule aligns perpendicularly to the light polarization, $\beta = \pi/2$. Studies by Wang et al. [57] and Shi et al. [58] show that alignment quality strongly depends on exposure to humidity and only slightly depends on the type of surface used for depositing BY. Our previous work showed that a longer light exposure produces a stronger in-plane (azimuthal) surface anchoring [40]. However, the effect of other important factors remained unexplored.

This study investigates the effect of photopatterning conditions, such as BY layer thickness ($b$), light intensity ($I$), irradiation dose, and age of non-irradiated substrate, on the alignment quality and the strength of azimuthal anchoring expressed by the anchoring coefficient $W$ in the surface potential $\frac{1}{2}W\sin^2\alpha$, where $\alpha$ is the angle between the alignment direction imposed by the patterned BY layer and the actual director specifying the local orientation of the liquid crystal. The coefficient $W$ is determined by analyzing the splitting of integer disclinations into half-integer disclinations at photopatterned substrates [26,40]. We find that $b$ in the range 5–8 nm yields the strongest azimuthal anchoring. $W$ increases with the dose and with $I$ when the dose is fixed. BY-coated substrates, which are photopatterned within 15 days of substrate preparation (after BY spin coating and baking, but before irradiation), show no significant change in $W$. However, substrates aged for more than 15 days before irradiation exhibit a decline in $W$. Our study also demonstrates that photopatterned cells filled with a conventional nematic LC and sealed with epoxy glue preserve their alignment strength for about 4 weeks, but further aging of the filled cell leads to a reduction in $W$. The results facilitate the optimization of BY photoalignment for liquid crystal applications.

## 2. Materials and methods

### 2.1. *Cell preparations*

Indium tin oxide (ITO)-coated glass plates are sonicated in water with a small amount of detergent at 60 °C for 15 min. Although the studies do not require an application of an electric field, the choice of ITO-coated plates is justified by the fact that most applications of liquid crystals involve electro-optic effects, thus the presence of the ITO electrodes is often a necessity. The plates are rinsed with isopropanol, dried in an oven at 80 °C for 15 min, and exposed to UV in an ozone chamber for 15 min. The plates are spin-coated with a solution of azo-dye BY in N, N-dimethylformamide (DMF), Figure 1a (both purchased from Sigma Aldrich, St. Louis, MO, USA) at 3000 rpm for 30 s, and baked at 80 °C for 30 min. We use DMF solutions with various BY concentrations, 0.2, 0.4, 0.5, 0.6, 0.8, 1.0, 2.0, and 4.0 wt%, in order to vary the thickness $b$ of the resulting BY layer and to explore its effect on the anchoring strength. All other experiments are performed with substrates coated with a 0.5 wt% BY solution, resulting in $b = 7.6$ nm, which is in the optimal thickness range of 5–8 nm that yields the strongest $W$. To avoid the detrimental effects of humidity on BY alignment [57], we limit the relative humidity (RH) of the environment to less than 20% during the spin coating and baking, and to 20–35% during substrate storage, cell assembly, and photopatterning. During imaging, the sealed LC filled cells are kept at an RH of less than 50% and a constant temperature of 45 °C.

Cells are assembled from two glass substrates with the BY-coated surfaces facing each other; these are called BY-BY cells throughout the text. The gap is fixed using epoxy glue NOA 65 (Norland Products Inc., Jamesburg, NJ, USA), without spacers, to achieve a thickness of ~1 μm. The thickness $h$ of the gap between two plates is measured by an interferometric technique using a UV/VIS spectrometer Lambda 18 (Perkin Elmer, Waltham, MA, USA)

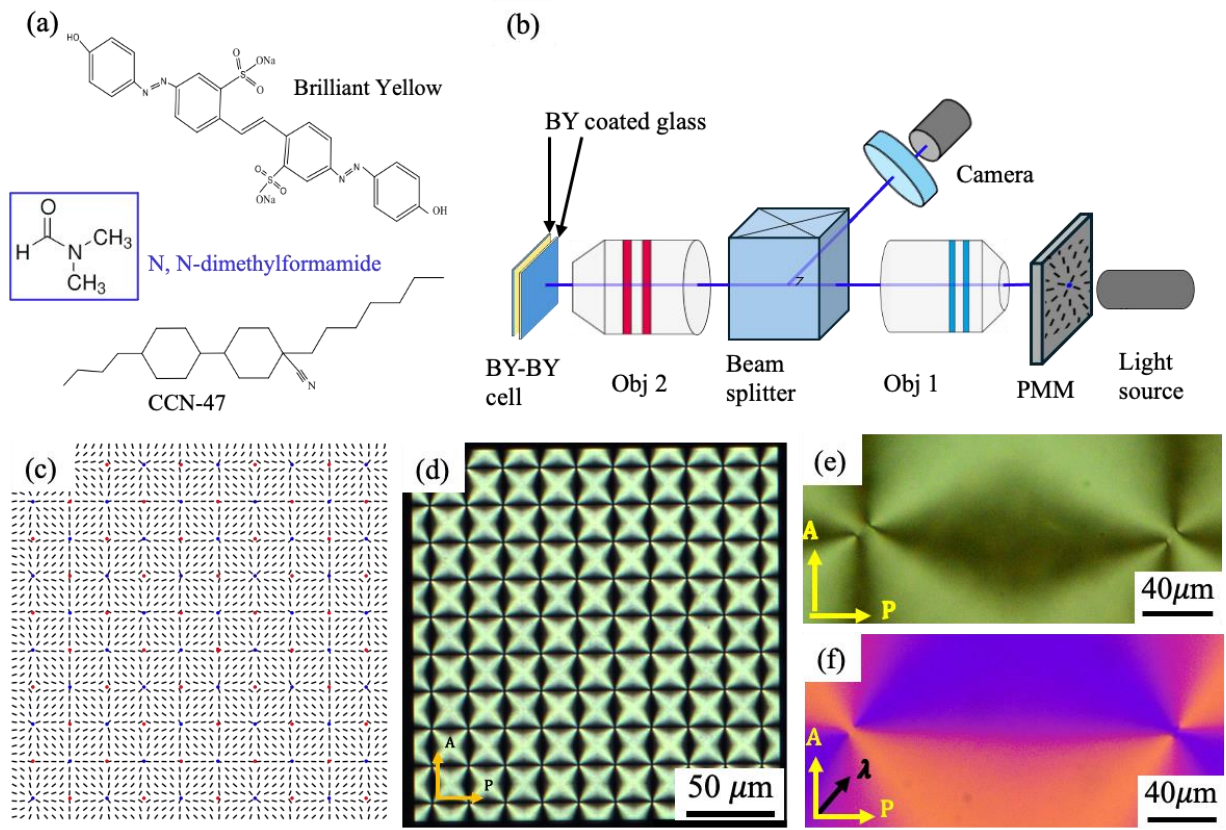

**Figure 1. Chemical structures of materials and photopatterned defect array.** (**a**) Photoresponsive azo-dye Brilliant Yellow, liquid crystal CCN-47, and solvent N, N-dimethylformamide. (**b**) Schematic of the photopatterning setup. Obj: objective, PMM: plasmonic metamask. (**c**) Director pattern of the defects array. The blue dots represent the cores of +1 defects, and the red dots represent the cores of −1 defects. (**d**) A polarizing optical microscope texture of the plasmonic meta mask with ±1 radial defect array. (**e**) Polarized optical microscope images of a portion of a photopatterned cell with a + 1 radial defect on the left and a − 1 defect on the right. The cell thickness is 1.1 μm. (**f**) The same polarizing microscopy with a full-wavelength optical compensator with the slow axis λ. P and A represent the polarizer and analyzer, respectively.

*2.2. Photopatterning*

We use the plasmonic metamask (PMM) technique introduced by Guo et al. [33,34] to pattern the substrates with an array of +1 and −1 defects (these defects should not be confused with "defects" induced by mishandling of the samples or by dust particles). PMM is an aluminum film with a thickness of 150 nm containing an array of nanoslits, each with a length of 220 nm and a width of 100 nm. When an unpolarized light beam passes through a nanoslit, the transmitted light becomes polarized along the short axis of the nanoslit. The degree of polarization of the transmitted light depends on the wavelength. Guo et al. [34] demonstrated experimentally and through numerical simulations that the polarization contrast ratio exceeds 7 dB for wavelengths ranging from 400 to 800 nm. The transmitted light beam irradiates the cells, as shown in Figure 1b. We use two types of cells in this work: BY-BY and BY-PS cells. The BY-PS cells are assembled with one BY-coated substrate and one polystyrene (PS)-coated substrate, with the coated surfaces facing each other. Preparation of the PS-coated substrates is explained later in the text, Section 3.1. During photoalignment, BY-PS cells are positioned so that the BY-coated substrate is closer to the light source. Irradiated light aligns azobenzene molecules perpendicularly to the light polarization direction. As a result, a desired pattern, which replicates the pattern of nanoslits in the PMM, is produced in the azo-dye layer. The BY molecule exhibits an absorption range of 350–500 nm, with a peak at 432 nm [49]. We use a light source EXFO X-Cite (Excelitas Technologies, Pittsburgh, PA, USA) with a wavelength range of (320–750) nm, which fully covers the absorption spectrum of BY. The light beam propagates along the normal to the PMM and the cell, Figure 1b. The light intensity $I$ is measured at the point of cell incidence using a power and energy meter console PM 100D (Thorlabs, Newton, NJ, USA).

A periodic square lattice of defects with strength +1 and −1 is designed by using a superposition rule for the in-plane director, $\hat{\mathbf{n}}_{BY} = (n_x, n_y, 0) = (\cos\varphi, \sin\varphi, 0)$, where $\varphi = \sum_{i=1}^{p}\sum_{j=1}^{q}(-1)^{i+j}\arctan\left(\frac{y-jb}{x-ia}\right)$, $x$ and $y$ are the Cartesian coordinates, $p$ and $q$ are the numbers of defects in rows and columns, respectively, $p = q = 10$; $a = b = 200$ μm is the distance between the defects along the $x$ and $y$ directions, respectively. The +1 defects in the pattern are of a radial type, so that the director around them experiences mostly splay, Figure 1c.

*2.3. Nematic material*

The photopatterned cells are filled with 4-butyl-4-heptyl-bicyclohexyl-4-carbononitrile (CCN-47), as shown in Figure 1a, by capillary action in the isotropic state at the temperature 70 °C. The material exhibits the following phase transitions upon heating: Smectic A 29.9 °C Nematic 58.5 °C Isotropic. After filling the cells, they are kept at 45 °C during the experiments using a Linkam hot stage (Linkam Scientific, Redhill, UK). At 45 °C, the elastic constants $K_1$ of splay and $K_3$ of bend for CCN-47 are equal, $K_1 = K_3 = K = 8$ pN [59,60], which allows one to use the superposition rule for the director field and to analyze the elastic properties of the patterns in the so-called one-constant approximation [61].

*2.4. Optical microscopy characterization*

Optical The optical textures of photopatterned LC cells are recorded using an Olympus BX51 polarized optical microscope (Olympus, Tokyo, Japan) equipped with a Basler (acA1920-155um) digital color camera (Basler, Ahrensburg, Germany), Figure 1e, f. A full-wavelength (530 nm) optical compensator is used to reconstruct the director field, Figure 1f. Regions where the director aligns parallel to the slow axis of the compensator exhibit a blue interference color, while areas where the director is perpendicular to the slow axis appear yellow. The left defect in Figure 1e, f is a + 1 radial defect, while the right one is a − 1 defect. The separation distance between half integer defects, $d$, is measured using the open-source software package Fiji/ImageJ, version 2.140/1.54f.

*2.5. Theoretical background*

Defects of strength ±1 tend to split into pairs of ±1/2 to reduce their elastic energy [61]. In the so-called one-constant approximation, the elastic repulsive potential of two +1/2 or two −1/2 defects is weakly dependent on their separation $d$: $F_E = -\frac{\pi K h}{2} \ln \frac{d}{2r_c}$, where $K$ is the average Frank elastic modulus, $h$ and $r_c$ are the cell thickness and the radius of the disclination core, respectively [26,61]. In a photopatterned cell, the separation of the defects is resisted by surface anchoring that tends to enforce the patterned ±1 defects. The elasticity-anchoring balance determines the equilibrium separation distance $d$ of the semi-integer cores. The surface anchoring energy of a patterned cell can be found by integrating the Rapini–Papoular potential, $F_S = 2\int_0^{2\pi}\int_0^d \frac{1}{2}W[1-(\hat{\mathbf{n}}_{BY} \cdot \hat{\mathbf{n}}_{LC})^2]\, r\, dr\, d\varphi$ which yields $F_S = 2\alpha W d^2$; here $\hat{\mathbf{n}}_{LC}$ is the actual director field of a split defect pair, $\hat{\mathbf{n}}_{BY}$ is the ideal radial pattern of the easy axis at the substrates, $\alpha \approx 0.184$ is a numerical coefficient, and the factor 2 reflects anchoring at both plates [26,40,61]. The equilibrium value of $d$ allows one to calculate the anchoring strength as $W = \pi K h/(8\alpha d^2)$. For cells assembled with one photopatterned surface and the second plate providing degenerate in-plane anchoring, $W = \pi K h/(4\alpha d^2)$.

We calculate $W$ by measuring $d$ and $h$ and using the known values of $K$ and $\alpha$. The resolution of the optical microscope is sufficient for precise measurements, as the experiments show that the parameter $d$ is higher than 3 μm even for the highest achieved anchoring. The distance between neighboring photopatterned +1 and −1 defects is set to be ~200 μm, which is sufficiently large to prevent any interactions between them. The experiments are designed with thin cells to ensure core splitting is the prevailing director structure as opposed to the escape in the third dimension [40,62]. Each data point for $d$ and $W$ represents the average value obtained from

50 defects of the same sign within the array, with the errors calculated as the standard deviation [63].

## 3. Results

3.1. *Effect of BY layer thickness on photoalignment*

The thickness $b$ of BY coatings affects the anchoring strength. Therefore, BY coatings with different thicknesses are produced by spin coating solutions with different BY concentrations. The thickness $b$ is measured using a digital holographic microscope (Lyncée Tec., Chino, CA, USA) in a reflection mode with a vertical resolution better than 1 nm. A portion of the BY coating is removed by wiping the substrate with water, followed by isopropanol using a cotton bud, exposing the glass surface to serve as the reference for thickness measurement, Figure 2a. After imaging the surface, the height difference between the coating and the glass substrate is measured using the open-source software package Fiji/ImageJ, Figure 2b. Thickness is measured at 10 different locations on the coated surface, and the average of these measurements is used as $b$. Changing the BY concentrations (0.2, 0.4, 0.5, 0.6, 0.8, 1.0, 2.0, and 4.0 wt%) in DMF results in different values of $b$, ranging from 3.2 to 80.9 nm, Figure 2c. The thickness $b$ was also verified by scratching the BY-coated glass with a sharp blade to create a groove; the depth $b$ of the groove is measured by a digital holographic microscope. The $b$ values obtained by the two methods are in good agreement with each other, being within the measurement error

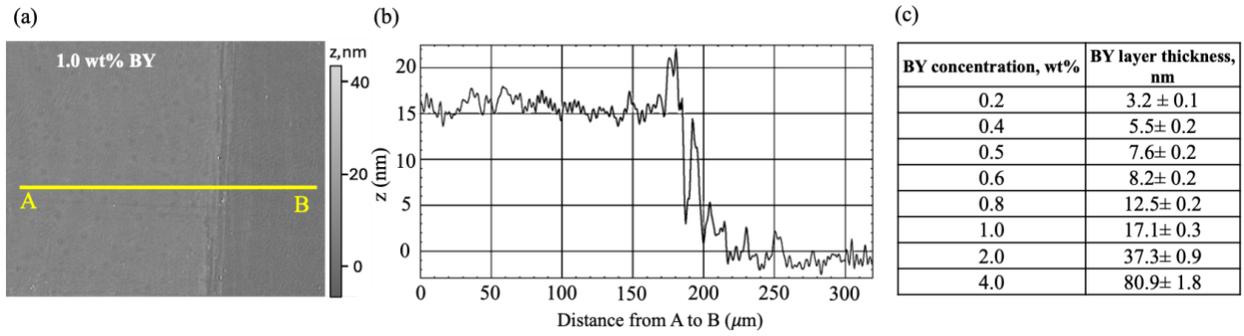

**Figure 2. Brilliant Yellow layer thickness measurement.** (**a**) An image of a 1.0 wt% BY-coated, partially wiped surface taken using a digital holographic microscope. The left side of the image shows the BY coating, while the right side shows the glass surface after wiping off the coating. (**b**) The surface profile plot along the line AB marked in (**a**). (**c**) BY layer thicknesses for different BY concentrations in the spin-coated solutions.

A larger $b$ reduces the intensity of light that reaches the second plate that is further away from the source due to absorption by the BY layer on the front plate. To explore the effect, we prepare the BY-PS cells with one BY-coated substrate as the front plate and a polystyrene (PS)-coated glass plate as the back substrate. During photoirradiation, the pattern is focused on the BY-coated front plate. PS-coated plates yield a negligibly weak azimuthal anchoring, $W_{PS} \sim 10^{-10}$ J/m$^2 \ll W$ [64]. Thus, it is the patterned substrate that dictates the director orientation and the separation distance between the defects. The anchoring coefficient for these BY-PS cells is calculated as $W = (\pi K h)/(4\alpha d^2)$. The PS-coated plates are prepared by spin coating (3000 rpm, 30 s) a solution of 0.5 wt% PS in chloroform (Sigma-Aldrich, >98%) on clean glass substrates. The plates are kept at 80 °C for 30 min to evaporate the solvent. All cell preparation steps and observations are conducted on the same day. The cells are photopatterned with $I = 5.50 \times 10^2$ Wm$^{-2}$ for 30 min.

The anchoring coefficient is measured for BY-PS cells with different $b$, Figure 3a–c. For a thin BY layer of 3.2 nm, $W$ is weak $\sim 0.2 \times 10^{-6}$ Jm$^{-2}$, Figure 3c. Anchoring increases sharply with $b$ and reaches a maximum of $0.98 \times 10^{-6}$ Jm$^{-2}$ for $b$ ranging from 5.5 nm to 8.2 nm, Figure 3c. For layers with $b = (12.5 - 37.3)$ nm, $W$ sharply decreases to around $10^{-7}$ Jm$^{-2}$ and remains constant for thicker BY layers, Figure 3c.

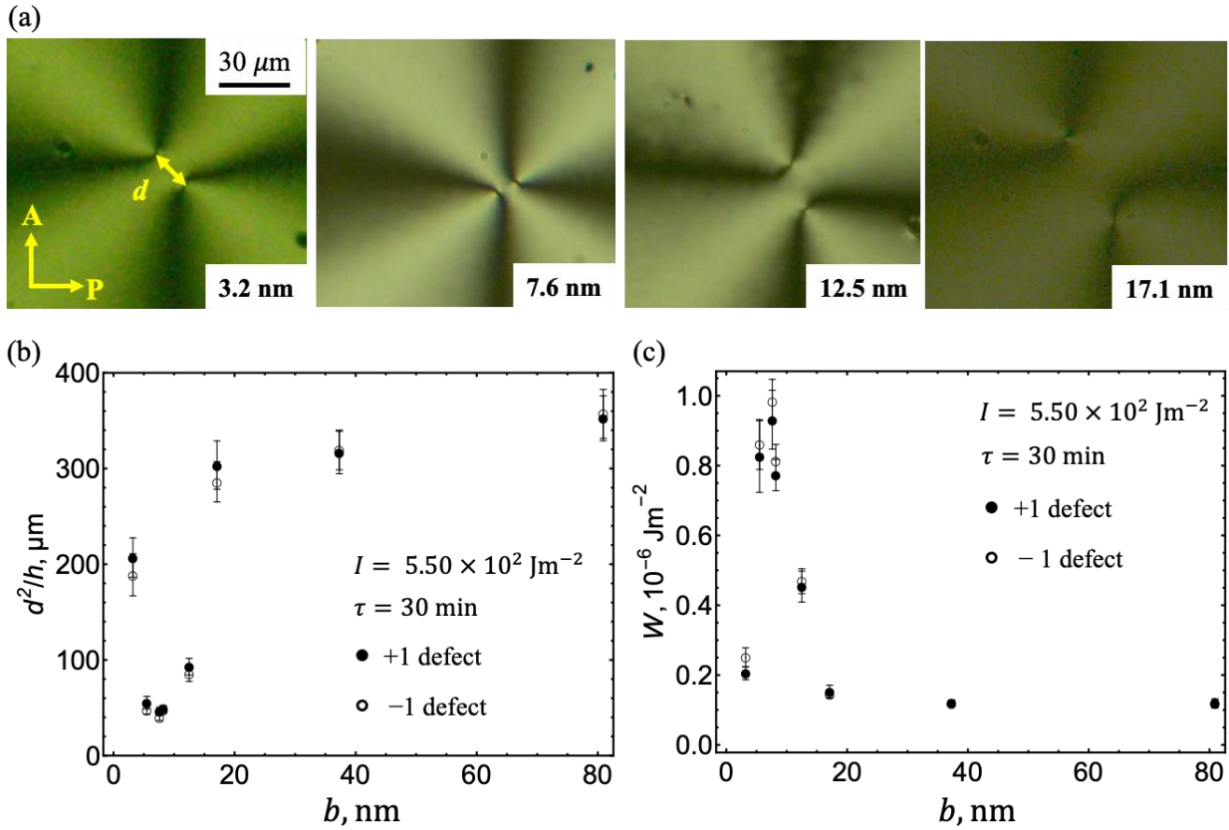

**Figure 3. Effect of Brilliant Yellow coating thickness on photoalignment in BY-PS cells.** (a) Optical microscopy textures of photopatterned cells with different BY layer thicknesses, $b$. $h = 2.2 \pm 0.2$ μm. Cells are photopatterned with $I = 5.50 \times 10^2$ Wm$^{-2}$ and a dose of $9.90 \times 10^5$ Jm$^{-2}$, corresponding to an irradiation time $\tau = 30$ min. Cell preparation and observation steps are performed on the same day. (b) $d^2/h$, and (c) $W$, as functions of $b$. Each data point represents

the average value obtained from 50 defects of the same sign within the array, with the errors calculated as the standard deviation.

The weak $W$ at the thinnest BY coatings ($h = 3.2$ nm) is most likely caused by the surface roughness of the glass, ITO, and the coating itself. As clear from Figure 2b, the thickness 3.2 nm is within the range of the measured variations of the coating's surface, which implies that in some places there might not be enough BY molecules. Another potential reason for weak anchoring at thin coatings is a formation of hydrogen bonds between the BY molecules and the hydrophilic UV/O$_3$-treated ITO surface. As noted by Wang et al. [57], BY films of 3 nm thickness on a hydrophilic substrate of polyvinyl alcohol (PVA) demonstrated a low degree of orientational order as compared to thicker films. The effect was attributed to the formation of hydrogen bonds between the BY molecules and the hydroxide OH group of PVA which hinders the *trans-cis* isomerization of BY molecules needed for light-induced alignment [57,65].

The weak $W$ at thick BY layers is attributed to weaker light intensity received by BY molecules at the interface with the LC, due to absorption by BY along the light path. The light transmittance through BY layers of varying thicknesses is measured at wavelength 410 nm using a UV/VIS spectrometer Lambda 18 (Perkin Elmer, Waltham, MA, USA). The data show 90% of incident light is transmitted by the BY layers when $b \leq 6$ nm, but the transmission is reduced to 22% for $b = 80.9$ nm, Figure 4b. As a result, the light intensity received by BY molecules on the surface that will meet the LC is substantially reduced, Figure 4a,b. As the effective intensity decreases, the effective dose also decreases; both factors yield a weaker $W$. Since BY layers with thicknesses less than 8 nm transmit over 85% of the incident light, and the thickness of 5–8 nm results in stronger azimuthal anchoring, a 7.6 nm thick BY-coating layer (spin coated with a 0.5 wt% BY solution) is used for the rest of experiments, presented in Figures 5–7. For a BY-BY cell, each with a 7.6 nm thick BY layer, the substrate that is further from the light source receives 87% of the incident light intensity during photoalignment. In such a cell, both BY surfaces that will later meet the LC are aligned with the same light intensity, 87% of the incident light. As shown in Figure 3c, this intensity is sufficient to achieve the strongest anchoring. Therefore, all subsequent experiments are performed using BY-BY cells assembled with two BY-coated glass substrates.

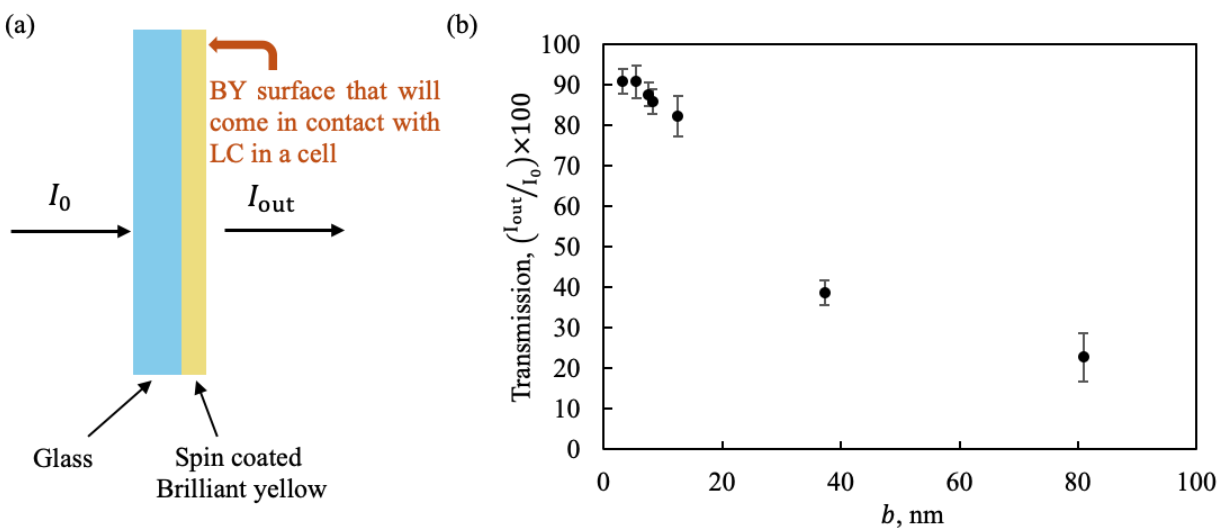

**Figure 4. Transmittance measurements of different Brilliant Yellow layer thicknesses.** (**a**) Schematic illustration of BY-coated glass. Here, $I_0$ is the incident light intensity and $I_{out}$ is the intensity of light transmitted through BY layer. (**b**) Transmission as a function of BY coating thickness. The wavelength of light is 410 nm.

*3.2. Effect of light dose and intensity on photoalignment*

The BY-BY cells are photopatterned with four different light irradiation doses. The dose is defined as a product of irradiation time ($\tau$) and the light intensity $I$. $\tau$ is changed to vary the dose while $I$ is kept constant. When studying the effect of $I$, the dose is kept constant by varying $\tau$. The cell preparation and characterization are completed within a maximum of two consecutive days to minimize the effects of aging.

Increasing the dose from $1.65 \times 10^5$ Jm$^{-2}$ to $9.90 \times 10^5$ Jm$^{-2}$ results in an increase in $W$, Figure 5(a–d rows) and 5e; $W$ saturates at doses higher than $4.6 \times 10^5$ Jm$^{-2}$. This behavior is consistent with the previous study on the effect of photopatterning time on $W$ [40].

Increasing $I$ from $1.25 \times 10^2$ Wm$^{-2}$ to $7.25 \times 10^2$ Wm$^{-2}$ at a fixed dose increases $W$ and creates better alignment, Figure 5(i–v columns) and 5f. It is notable that although dose and $I$ are interrelated, both must be set properly to achieve a high-quality patterning. For instance, photopatterning with a high dose of $9.90 \times 10^5$ Jm$^{-2}$ and a low $I$ of $1.25 \times 10^2$ Wm$^{-2}$ results in a poor alignment (Figure 5d.i), and a weak $W = 0.36 \pm 0.02 \times 10^{-6}$ Jm$^{-2}$, Figure 5e,f. However, a moderate dose of $2.25 \times 10^5$ Jm$^{-2}$ at $I > 5.50 \times 10^2$ Wm$^{-2}$ results in a better alignment with $W \geq 0.4 \times 10^{-6}$ Jm$^{-2}$. Using $W$ values, one can establish criteria for setting irradiation conditions to achieve a good alignment. The results also show that $W$ can be tuned by adjusting the dose and $I$.

At a constant dose, photopatterning with high $I$ for a short $\tau$ is more efficient than low intensity irradiation over a prolonged $\tau$ (compare columns i and v in Figure 5), Figure 5f. This indicates that the total number of photons is not the single decisive factor defining the anchoring strength. The intensity-dependent behavior of $W$ suggests that photoisomerization occurs as a collective process, where the isomerization of individual BY molecules depends on the isomerization probability of neighboring molecules. Schönhoff et al. [66] observed a similar dependency of molecular photoreorientation on intensity for 4-(4′-N-octadecylamino)phenylazocyanobenzene (amino azobenzene) films irradiated by polarized light with varying intensity but a constant dose and concluded that photoreorientation is a pronounced collective effect.

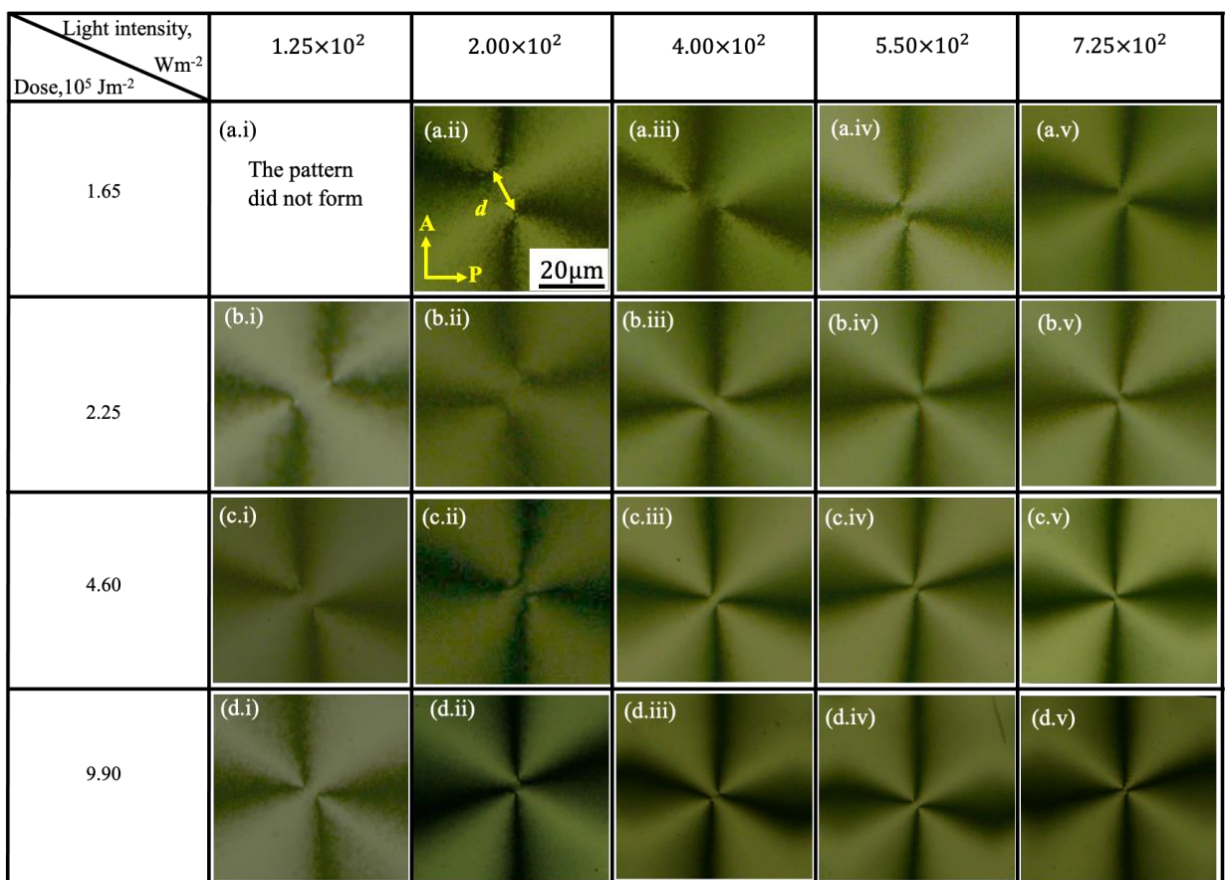

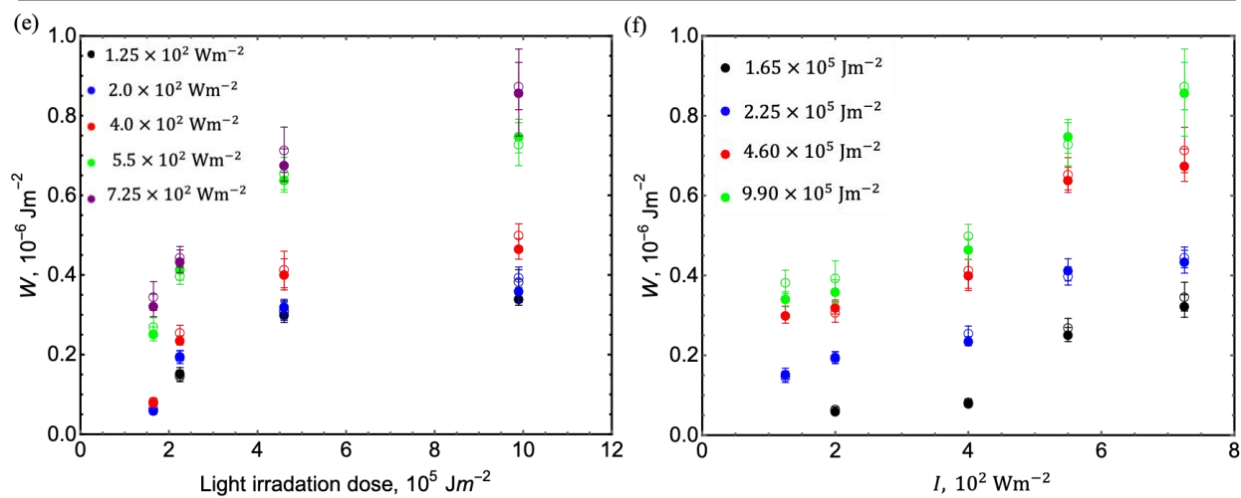

**Figure 5. Effect of dose and light intensity on photoalignment in BY-BY cells.** (**a**–**d**) Optical microscopy textures of photopatterned cells at different irradiation doses and light intensities. The images show a pair of +1/2 defects split from a photopatterned +1 defect. $h = 1.3 \pm 0.3$ μm and $b = 7.6$ nm. Cell preparation and characterization are performed within two consecutive days. (**e**)

Azimuthal surface anchoring coefficient, $W$, as a function of dose for various light intensities. (**f**) $W$ as a function of $I$ for different doses. Colors indicate different intensities and doses in (**e**) and (**f**), respectively. Filled circles represent +1 defects, and open circles represent –1 defects. Each data point for $W$ represents the average value obtained from 50 defects in the array, with the errors calculated as the standard deviation.

*3.3. Effect of aging of non-irradiated BY coatings.*

The glass substrates, which were spin-coated with BY and then baked, are stored in a Humidity and Temperature-Controlled Cabinet (SIRUI HC series) for up to 310 days at relative humidity (RH) 25-35%, and 23 °C. Then BY-BY cells are assembled using two aged, non-irradiated BY-coated plates. After a predetermined aging time, the BY coatings are photopatterned with an $I$ of $5.50 \times 10^2$ Wm$^{-2}$ for 30 min. The cells are filled with the nematic material, sealed and analyzed under the optical microscope

Aging of the non-irradiated BY layer impacts photopatterning in two phases. For BY layers irradiated within 15 days after the layers were spin-coated, dried, and baked, $W$ remains constant at $\sim(0.88 \pm 0.01) \times 10^{-6}$ Jm$^{-2}$, Figure 6. This suggests that BY-coated substrates can be safely stored at controlled humidity conditions (at RH 25–35%) for around two weeks without a noticeable effect on their photopatterning quality. However, for older BY layers, $W$ decreases continuously with age and reduces to $0.08 \times 10^{-6}$ Jm$^{-2}$ for substrates photopatterned 310 days after preparation, Figure 6. A possible reason could be the water absorption by the BY layers during storage. As already noted, we control the RH of the environment at less than 20% during the spin coating and baking, but during the storage, the RH is at higher levels 20–35%. Absorption of water during the prolonged storage can result in aggregation of BY molecules into J-structures, a process called by Shi et al. [58] "hydrogen-bond-assisted self-assembly of BY molecules with water molecule insertion". The molecules in J-aggregates are less likely to undergo an efficient trans-to-cis isomerization. Accumulation of water at the substrate–BY interface with a suppression of isomerization through hydrogen bonds might also contribute to the effect of aging.

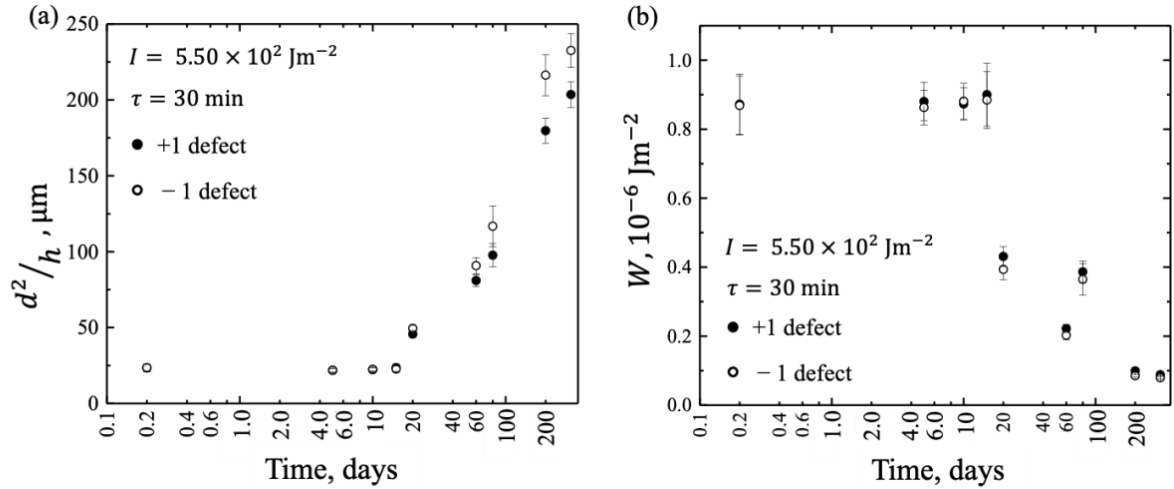

**Figure 6. Effect of non-irradiated BY coated layer aging on photoalignment in BY-BY cells.** (a) $d^2/h$, and (b) $W$, as a function of non-irradiated BY layer aging time. Each data point represents the average value obtained from 50, +1 and 50, −1 defects within the array, with the errors calculated as the standard deviation. $h = 1.1 \pm 0.2$ μm and $b = 7.6$ nm. Cell assembly, photopatterning, and characterization are performed on the same day. Cells are photopatterned with $I = 5.50 \times 10^2$ Wm$^{-2}$ and dose of $9.90 \times 10^5$ Jm$^{-2}$.

*3.4. Surface patterning stability of aged LC-filled cells.*

A photopatterned BY-BY cell is filled with CCN-47, and the edges of the cell are sealed with epoxy glue after the filling. Photopatterning is performed with $I = 5.50 \times 10^2$ Wm$^{-2}$ for 30 min. The LC-filled BY-BY cell is maintained in an environment with RH <50% at 45 °C for 72 days. While maintaining the temperature at 45 °C to stabilize the nematic phase, optical textures of the cell are recorded over time as the cell ages.

The distance $d$ remains at around $10.4 \pm 0.3$ µm for 26 days, corresponding to $W = (0.82 \pm 0.01) \times 10^{-6}$ Jm$^{-2}$, Figure 7a,b. Maintaining the cell for a longer period (around 72 days) results in a slight decrease in $W$ to $0.76 \times 10^{-6}$ Jm$^{-2}$, Figure 7b. This result suggests that filling the photopatterned cell with LC and sealing it helps to preserve the quality of photopatterning. The observed decline could be due to the gradual dissolution of LC into the BY coating [67,68] or the dissolving of epoxy glue into the LC over time [69].

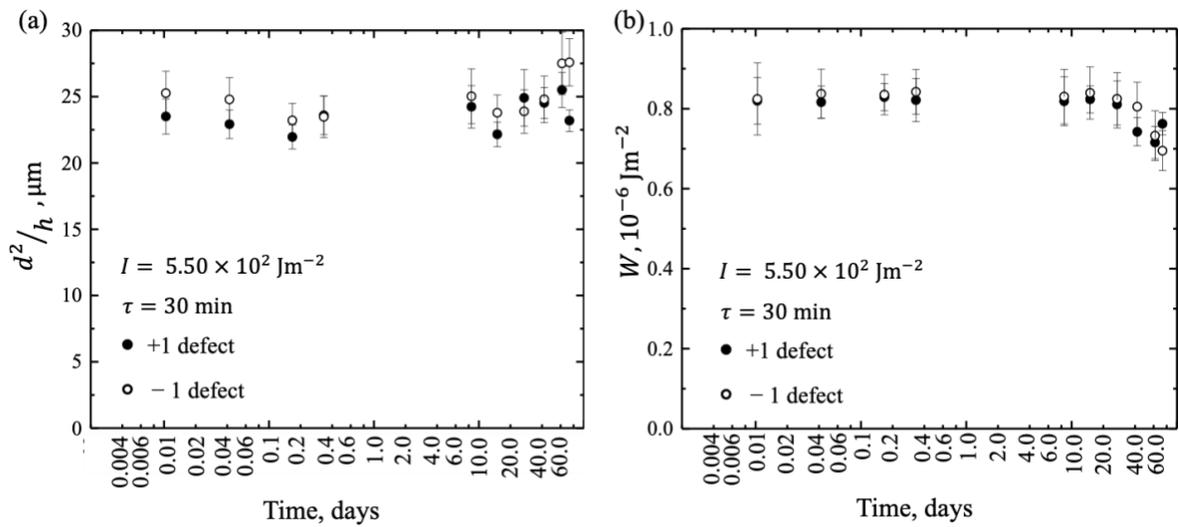

**Figure 7. Surface patterning stability of aged LC-filled BY-BY cells.** (a) $d^2/h$, and (b) $W$, as functions of the age of the LC filled cell. Each data point represents the average value obtained from 50 defects of the same sign within the array, with the errors calculated as the standard deviation. $h = 1.5 \pm 0.02$ µm and $b = 7.6$ nm. The cell is photopatterned with $I = 5.50 \times 10^2$ Wm$^{-2}$ and a dose of $9.90 \times 10^5$ Jm$^{-2}$. Cell preparation steps are performed on the same day and observation is done as the cell ages.

## 4. Conclusion

We have demonstrated how various photopatterning conditions, including the thickness $b$ of the azobenzene alignment layers the intensity $I$ of the light used for photopatterning, the irradiation dose, and substrate aging, affect the anchoring strength of a photopatterned nematic LC. Our results show that BY layers of thickness $b = 5$–$8$ nm produce the strongest anchoring coefficient $W$. Moreover, $W$ increases with both irradiation dose and $I$. Also, aging of non-irradiated substrates beyond 15 days significantly reduces $W$. However, if the cell is filled with a LC immediately after photopatterning of the BY layer, $W$ remains constant for up to 4 weeks. This work provides practical strategies for enhancing the azimuthal strength of the photopatterned anchoring of nematics. The results offer guidelines for optimizing BY photoalignment parameters and storage.

The method to measure the azimuthal anchoring coefficient described in this paper is based on the properties of topological defects produced by photopatterning; it does not require one to use a second plate, for example, a rubbed polyimide plate, with an anchoring much stronger than the photoalignment anchoring [47,51,70]. It also does not require one to use any external fields [71-73] or to prepare wedge samples of varying thickness [74-76]. Within a broader prospectus, our approach to measuring the azimuthal anchoring coefficient can be extended to other photoalignment materials, such as SD-1 that does not feature absorption peaks in the visible spectral range [49] and thus might be better suited for applications that require a long-term stability.

**Author Contributions**

N.P.H. did the experiments, analyzed and discussed the data, and contributed to the writing, M.R. discussed the data and contributed to the writing, O.D.L. contributed to the writing and directed the research. All authors have read and agreed to the published version of the manuscript.

**Funding**

The work is supported by NSF grant DMR-2215191.

**Data availability statement**

The datasets generated during and/or analyzed during the current study are available from the corresponding author upon reasonable request.

**Conflicts of Interest**

The authors declare no conflict of interest.